\documentclass[conference]{IEEEtran}

\IEEEoverridecommandlockouts

\usepackage{array}
\usepackage{url}
\usepackage{subfigure}
\usepackage{amsmath,amssymb,stmaryrd}
\usepackage{multirow}
\usepackage{amsmath}
\usepackage{amstext}
\usepackage{cite}
\usepackage{graphicx}
\usepackage{tikz}
\usepackage{eqparbox}
\usepackage[utf8]{inputenc}
\usepackage{url}
\newcommand{\pe}{P_{\textrm e}}
\newcommand{\erfc}{\textrm{erfc}}
\newcommand{\sigs}{\sigma^2}

\newcommand{\ptarget}{P_{\textrm{trgt}}}

\tikzstyle{boss}=[rectangle, minimum width=3cm, minimum height=1cm, text centered, draw=black]
\tikzstyle{block}=[rectangle, minimum width=1cm, minimum height=1cm, text centered, draw=black]
\tikzstyle{ldpc}=[rectangle, minimum width=5cm, minimum height=2cm, text centered, draw=black]

\begin{document}

\title{Design of Polar Code Lattices of Small Dimension}

\author{\IEEEauthorblockN{Obed Rhesa Ludwiniananda$^1$, Ning Liu$^2$, Khoirul Anwar$^1$, and Brian M.~Kurkoski$^2$}
	\IEEEauthorblockA{$^1$Telkom University, Bandung, Indonesia \\
	$^2$Japan Advanced Institute of Science and Technology, Nomi, Ishikawa, Japan\\
		}
\thanks{This work was supported by JSPS Kakenhi Grant Number JP 19H02137. This work is also the output of the ASEAN IVO project, PATRIOT-41R-Net,  financially supported by NICT, Japan.  
}
}

\maketitle

\begin{abstract}
Polar code lattices are formed from binary polar codes using Construction D.  In this paper, we propose a design technique for finite-dimension polar code lattices. The dimension $n$ and target probability of decoding error are parameters for this design.  To select the rates of the Construction D component codes, rather than using the capacity as in past work, we use the explicit finite-length properties of the polar code.  Under successive cancellation decoding, density evolution allows choosing code rates that satisfy the equal error probability rule.   At an error-rate of $10^{-4}$, a dimension $n=128$ polar code lattice achieves a VNR of 2.5 dB, within 0.2 dB of the best-known BCH code lattice, but with significantly lower decoding complexity.  
\end{abstract}

\section{Introduction}

An $n$-dimensional lattice $\Lambda$ is a discrete additive subgroup of 
the $n$-dimensional Euclidean space $\mathbb{R}^n$.  In communications applications, lattices can provide shaping gain, and are integral to certain Gaussian network coding approaches including compute-forward relaying \cite{NazerG11} and integer-forcing MIMO \cite{zhan14IF}.  Low-dimension lattices with $n \leq 24$ have been well understood for some time \cite{conway2013sphere}, and recently results on lattices with dimension $n \geq 1000$ have appeared \cite{liu-com19} \cite{daSilva-it19}.  However, there has been relatively little study of good lattices with $24 < n < 1000$, which is relevant for low-latency communications.

Construction D builds lattices from two or more component codes, which are linear binary codes. Polar code lattices are appealing because the lattice inherits the good properties of the underlying polar codes. Liu \textit{et al.~}showed that polar code lattices are AWGN good, and can achieve the capacity of the AWGN channel \cite{liu-com19}.  

Polar codes have flexibility in rate selection, which is important for good lattice design. As far as we know, the only existing design of a finite-dimension lattice is by Liu et al.; a dimension $n=1024$ polar code lattice \cite{liu-com19}.  The design was guided by channel capacity, but the actual design differed noticeably from what would be predicted by capacity. In high dimension, it is reasonable to use channel capacity as design guidance, but this breaks down when considering small and medium-dimension lattices.  Questions still remain about the best way to design finite dimension polar code lattices. 

As a multi-level construction, Construction D lattices can be designed using the decoder error rates for the component codes, under the equal error probability rule \cite{Wachsmann-it99}. This has also been used to design Construction D' lattices based on LDPC codes \cite{daSilva-it19}; the shortcoming is that Monte Carlo simulations are slow and time-consuming. However, for binary polar codes with successive cancellation decoding, decoder error rates can be obtained using density evolution \cite{Mori-comlett09}. 

This paper contributes a design technique for polar code lattices of finite dimension. As a Construction D lattice, the challenge is to select the rates of the component codes that give the best lattice properties. Rather than using the capacity, we use the explicit finite-length code properties in the design.  For a code of block length\footnote{The block length of a code and the dimension of a lattice are both denoted $n$; for any design the block length of code is the dimension of the lattice.} $n$, we define $\rho$ as the code rate such that its decoder achieves a given target error rate, as a function of the channel noise $\sigs$. This allows systematic and efficient design of polar code lattices for a target error rate. Under successive cancellation decoding, the $\rho$ function is found efficiently using density evolution. The function $\rho$, based on finite-length code properties, has an S-shape which is characteristic of a channel capacity curve.

Under successive cancellation decoding, a dimension $n=128$ and $n=256$ polar code lattice achieves VNR of $3.25$ dB and $3.0$ dB respectively, at WER of $10^{-4}$.  Under successive cancellation list decoding, a dimension $n=128$ polar code lattice achieves VNR of $2.5$ dB.  This is within $0.2$ dB of the best-known $n=128$ BCH code lattice \cite{matsumine2018bch}, and SCL decoding complexity is significantly lower than OSD decoding.

\section{Construction D Lattices}
\label{section:ConstructionD}

\subsection{Construction D Lattice}

Construction D is a type of lattice construction \cite{conway2013sphere} \cite{Barnes-cjm83}. Construction D uses multiple nested binary codes to produce a multilevel construction \cite{imai1977new}. 
Let $C_0 \subseteq C_1 \subseteq \cdots \subseteq C_{a-1} \subseteq C_a = \mathbb{F}_{2}^{n}$ be nested binary linear codes with generator matrices $\mathbf{\tilde{G}}_0 , {\cdots}, \mathbf{\tilde{G}}_{a-1}, \mathbf{\tilde{G}}$, respectively.  Code $C_i$ is an $(n,k_i)$ binary with code rate $R_i = k_i/n$, and the code nesting gives:
\begin{align*}
R_0 \leq & R_1 \leq \cdots \leq R_a \\
k_0 \leq & k_1 \leq \cdots \leq k_a 
\end{align*}

\begin{figure}
				\centering
				\includegraphics[width=0.48\textwidth]{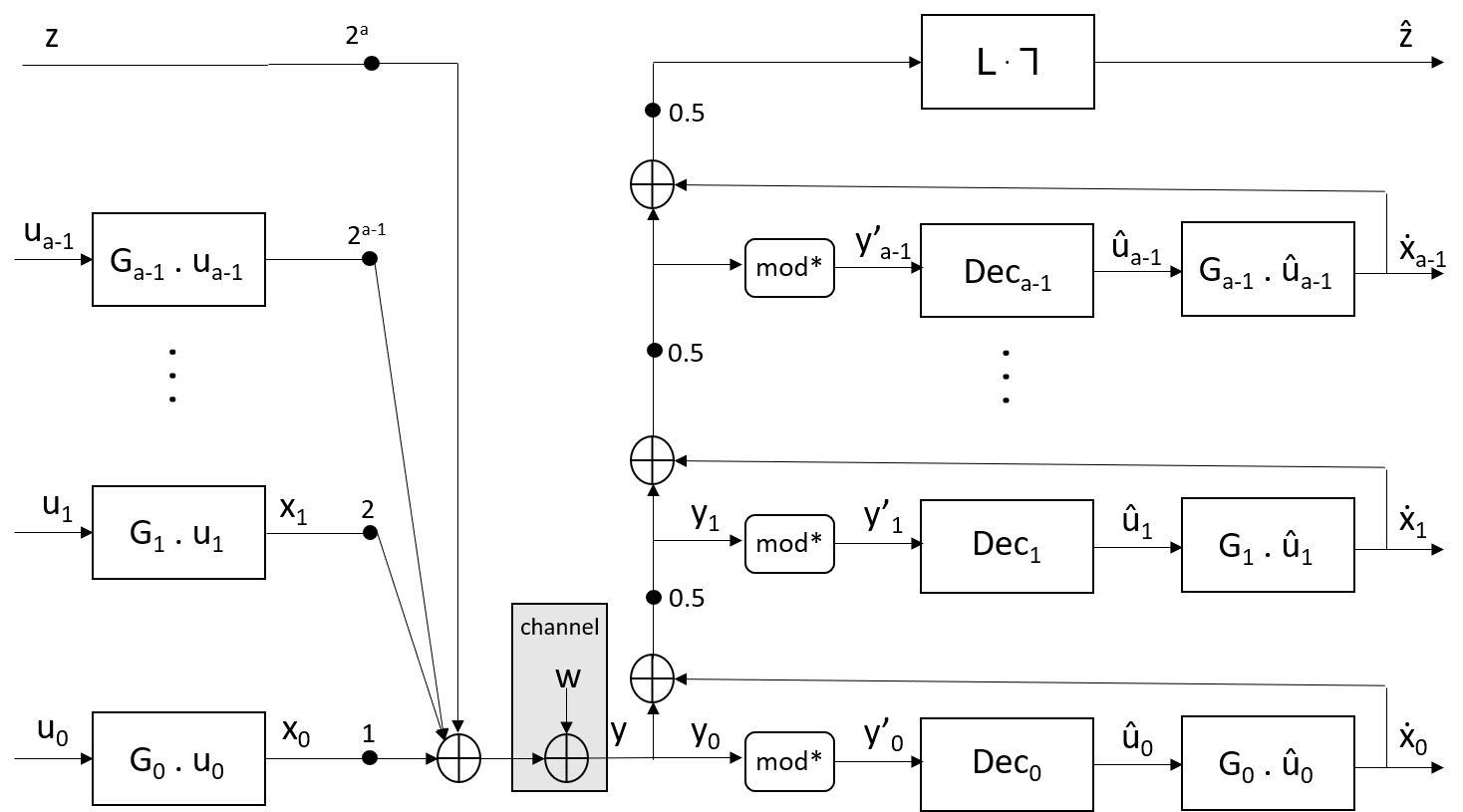}
				
				\caption{Encoder, channel and multistage decoder for Construction D} \label{fig:consD}
			\end{figure}

Construction D lattice $\Lambda$ consists of all vectors of the form
\begin{eqnarray}
\mathbf{x} = \sum_{i=0}^{a-1} 2^i \widetilde{\mathbf{G}}_i \cdot \mathbf{u}_i + 2^a \widetilde{\mathbf{G}}  \cdot \mathbf{z}, \label{eqn:constructionDdecomposition}
\end{eqnarray}
where $\mathbf{z} \in \mathbb{Z}^n$ and 
$\mathbf{u}_i = (u_{1,i},u_{2,i\cdots},u_{ki,i})^t, i \in  0,1,...,a-1$ and $u_{j,i} \in \{0,1\}$ are information bits; operations are over the real numbers and not the binary field. The full Construction D system of encoder, channel and decoder is depicted in Fig.~\ref{fig:consD}.

\subsection{Construction D Generator Matrix} 

The Construction D generator matrix is constructed using $\widetilde{\mathbf{G}}$, a  specific basis for $\mathbb F_2^n$, and $k_0,k_1,...,k_{a-1}$. The Construction D generator matrix $\mathbf{G}$ is given by:
\begin{eqnarray}
\mathbf{G} = \widetilde{\mathbf{G}} \cdot \mathbf{D}^{-1},
\end{eqnarray}
where $\mathbf{D}$ is a diagonal matrix with diagonal entries $d_{ii}$ :
\begin{eqnarray}
d_{ii} = 2^{-k} \;\textrm{for} \, r_{k-1} \leq i \leq r_k  
\end{eqnarray}
with $k = {0,1,...,a}$. 

\subsection{VNR and Channel Model}

For Construction D, the lattice volume $V(\Lambda) = | \det(\mathbf G)|$ is
\begin{eqnarray}
V(\Lambda) = 2^{an-n\sum_{i=0}^{a-1} R_i}.
\end{eqnarray} 
The unconstrained power channel is used in this paper, where an arbitrary $\mathbf x \in \Lambda$ is transmitted over an AWGN channel with noise power $\sigma^2$, and received as $\mathbf y$.  This allows evaluating the coding properties of the lattice without considering a specific shaping constellation.  Instead of a transmit power constraint, the lattice is constrained by the volume of the Voronoi region.  The volume-to-noise ratio (VNR) is
\begin{align}
    \textrm{VNR} &= \frac{V(\Lambda)^{2/n}}{2 \pi e \sigma^2}.
\end{align}
so that $\textrm{VNR}$ is the distance to the Poltyrev limit. 

\subsection{Decoding for Construction D}

Construction D decoding uses successive cancellation decoding, decoding $C_0$ first, then $C_1,C_2,\ldots$ in order (not to be confused with successive cancellation decoding of polar codes). Before the component code decoding, a modulo operation is performed to preserve distances to (0,1)
\begin{eqnarray}
\textrm{mod}^*(y_i) = \begin{vmatrix}
\textrm{mod}_2(y_i+1)-1
\end{vmatrix}
\end{eqnarray}
The result of modulo operation is the input to a binary polar code decoder.  The decoder produces an estimate of the information bits $\mathbf{\hat{u}}_i$ and to obtain $\mathbf{\hat{x}}_i$ use re-encoding with generator matrix
$\mathbf{\hat{x}}_i = \mathbf{G}_i \cdot \mathbf{\hat{u}}_i$
As at the encoder side, multiplication is over the reals. This estimate $\mathbf{\hat{x}}_i$ is subtracted from the input, and this is divided by 2,
$y_{i+1} = \frac{y_i-\hat{x}_i}{2}$
The estimated lattice point is
$\mathbf{\hat{x}} = \mathbf{\hat{x}}_1 + 2\mathbf{\hat{x}}_2 + \cdots + 2^{a-1}\mathbf{\hat{x}}_{a-1} + 2^a\mathbf{\hat{z}}$.

\section{Polar Code Lattices}
\label{section:polarCodeLatttices}

\subsection{Polar Codes and Density Evolution}

Polar codes were introduced by Arıkan \cite{arikan2009polar}. They provably achieve the symmetric capacity of binary input discrete memoryless channels with a low complexity decoder. Polar codes of block length $n$ have two types of bits, $k$ information bits $\mathcal I \subset \{1,2,\ldots,n\}$ and $n-k$ frozen bits; the rate is $R= k/n$. Successive cancellation (SC) decoding makes an estimate for bit position $u_j$ using hard decisions or frozen bits $\widehat u_1^{j-1}$ and the recursive LLR computation:
\begin{align*}
    L_n^{(2j-1)}(y_1^n, \widehat u_1^{2j-2}) &= 2 \tanh^{-1}\left( \tanh( \frac{\alpha_{n,j}} 2) \cdot \tanh ( \frac{\beta_{n,j}} 2) \right)
    \\
    L_n^{(2j)}(y_1^n, \widehat u_1^{2j-1}) &= (-1)^{\widehat u_{2j - 1}} \alpha_{n,j} +  \beta_{n,j}
\end{align*}
where
\begin{align*}
    \alpha_{n,j} &= L^{(j)}_{n/2} (y_1^{n/2}, \widehat u_{1,\textrm{even}}^{2j-2} + \widehat u_{1,\textrm{odd}}^{2j-2} ),
    \\
    \beta_{n,j} &= L^{(j)}_{n/2}(y_{n/2+1}^{n}, \widehat u_{1,\textrm{even}}^{2j-2} ).
\end{align*}

Polar code design assigns the position of information bits and frozen bits to indices. Mori and Tanaka described how to design polar codes using density evolution \cite{Mori-comlett09}, under SC decoding.  Under the assumption of a symmetric channel, analysis of the all-zeros codeword is sufficient.  Let the probability density function of the memoryless channel LLR message be $a_1^{(1)}(x)$.  Then the densities may be calculated as:
\begin{align*}
    a_{2n}^{(2j)}(x) & = (a_n^{(j)} \star a_n^{(j)})(x)
    \\
    a_{2n}^{(2j-1)}(x) & = (a_n^{(j)} \boxast a_n^{(j)})(x),
\end{align*}
for $j=1,2,\ldots, n$ where $\star$ is standard convolution for the variable node and $\boxast$ is specific check node convolution operation \cite[p.~181]{Richardson-2008} For a block length $n$ polar code, the distribution $a_n^{(j)}(x) = \Pr( L_j = x | \widehat u_{1}^{j-1} = 0)$ is used to make a hard decision in position $j$.  The probability of error in position $j$ given positions $1$ to $j-1$ are correct is:
\begin{align}
    p_j = 
    \int_{-\infty}^0 a_n^{(j)}(x) dx. \label{eqn:pi} 
\end{align}
The information bits $\mathcal I$ are selected to be the $k$ positions with the $k$ smallest values of $p_j$.

\subsection{Polar Code Lattices}

Polar code lattices are formed using $a \geq 2$ polar codes $C_0, \ldots, C_{a-1}$ \cite{liu-com19}. The polar code generator matrices satisfy the requirement for Construction D if the polar codes satisfy the nesting condition, $C_0 \subseteq C_1 \subseteq \cdots \subseteq C_a$.  For each polar code $C_i$ the information set is $\mathcal I_i$.  Then, a sequence of polar codes can be used to form a polar code lattice if $\mathcal I_0 \subseteq \mathcal I_1 \subseteq \cdots \subseteq \mathcal I_a$.  Thus, the information set structure can naturally form the subcodes needed for Construction D.

\emph{Example 1:} Consider a basis for $\mathbb{F}_2^4$ given by $\widetilde{\mathbf{G}}$, and nested binary codes with $k_0 = 2$, $k_1 = 3$, with generator matrices $\widetilde{\mathbf{G}}_0$ and $\widetilde{\mathbf{G}}_1$, given as:
\begin{eqnarray}
\widetilde{\mathbf{G}} &=& (\mathbf{G_m})^t \nonumber \\
\widetilde{\mathbf{G}} &=& \begin{bmatrix}
1 & 1 & 1 & 1\\ 
0 & 1 & 0 & 1\\ 
0 & 0 & 1 & 1\\ 
0 & 0 & 0 & 1
\end{bmatrix} \\ \nonumber
\end{eqnarray}

\begin{eqnarray*}
\widetilde{\mathbf{G}}_0 = \begin{bmatrix}
  1 & 1\\ 
  0 & 1\\ 
  1 & 1\\ 
  0 & 1
\end{bmatrix} \textrm{ and }
\widetilde{\mathbf{G}}_1 = \begin{bmatrix}
 1 & 1 & 1\\ 
 1 & 0 & 1\\ 
 0 & 1 & 1\\ 
 0 & 0 & 1
\end{bmatrix}.
\end{eqnarray*}
From Construction D definition with $a = 2$, a lattice point satisfies
\begin{eqnarray}
x = \widetilde{\mathbf{G}}_0 \cdot \mathbf{u}_0 + 2\widetilde{\mathbf{G}}_1	\cdot \mathbf{u}_1 + 4\widetilde{\mathbf{G}} \cdot \mathbf{z},
\end{eqnarray}
where $\mathbf{u}_0 = [u_{0,0},u_{0,1}]^t$ and $\mathbf{u}_1 = [u_{1,0},u_{1,1},u_{1,2}]^t$ are binary vectors and $\mathbf{z} \in \mathbb{Z}^4$.

\begin{figure}[t!]
            \begin{center}
                 \includegraphics[width=0.48\textwidth]{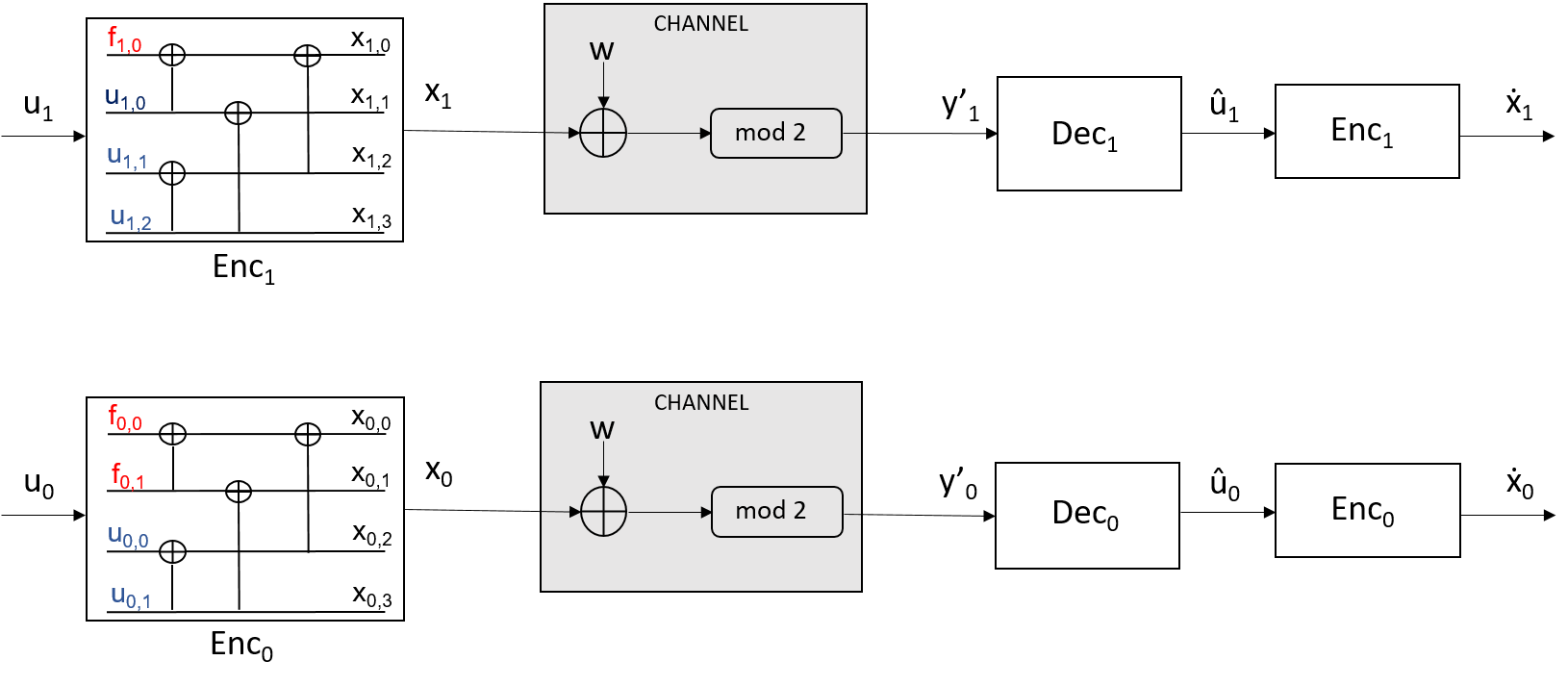}
                \caption{Equivalent encoder, channel and decoder for multilevel decoding of Construction D assuming each level is decoded successfully.}
                \label{fig:kfunction}
            \end{center}
        \end{figure}

\subsection{Equal Error Probability Rule Design Method }

Due to the multilievel structure of the decoder, level $i$ is decoded successfully only if level $i-1$ was decoding successfully.  Assuming successful decoding at all levels, each level can be seen as coding over independent channels; Fig.~\ref{fig:kfunction} shows $\textrm{Enc}_0$ and $\textrm{Enc}_1$ as an illustration of Example 1.  Each channel is a modulo AWGN channel called AMGN (additive modulo Gaussian noise) channel.  The channel noise variances are $\sigma^2, \sigma^2/4, \sigma^2/16 \ldots$ for level $i=0,1,2, \ldots$, respectively. 

A lattice decoder error is the event that $\widehat{\mathbf x} \neq \mathbf x$, and $\pe$ is the probability of this event (also, word error rate (WER) is $\pe$).  Let $\pe(C_i,\sigma^2)$ be the probability of decoder error for $C_i$ when used on the AMGN with noise variance $\sigma^2$. Then by the union bound:
\begin{align}
P_e \leq \pe(C_0,\sigs) + \pe(C_1,\frac{\sigs}4) + \cdots + \pe(C_a,\frac{\sigs}{4^a}) \label{eqn:unionBound}.
\end{align}

The equal error probability rule was given by Wachsmann et al.~for the design of multilevel codes \cite{Wachsmann-it99}, and Construction D lattices are a special case of a multilevel code.  Under the equal error probability rule, the codes $C_i$ are selected such that $\pe(C_i,\frac{\sigs}{4^i})$ are equal.

Using $n=1024$ and $a=2$, Liu et al. gave a polar code lattice design with $(R_0,\pe(C_0)) = (0.23, \frac 1 3 \cdot 10^{-5})$, $(R_1,\pe(C_1)) = (0.9, \frac 1 3 \cdot 10^{-5})$, and $(R_2,\pe(C_2)) = (1, \frac 1 3 \cdot 10^{-5})$.  This design achieved VNR of $2.34$ dB for $\pe=10^{-5}$ (and VNR $\approx 2.05$ dB for $\pe=10^{-4}$) \cite{liu-com19}.  These code rates differ noticeably from those predicted by a capacity-oriented design, due to capacity losses at finite length.

\section{Proposed Design Method}
\label{section:ProposedDesignMethod}

In this section we describe the proposed polar code lattice design, which explicitly uses the finite-length properties of polar codes with the equal error probability rule.

\subsection{$\rho$ Function for Binary Polar Codes}

Consider a binary polar code $C$ on the $\sigma^2$ AMGN channel and a target decoder error rate of $\ptarget$. As the code rate increases, the SNR $= 1 / \sigma^2$ needed to achieve a given word-error rate also increases.  Equivalently as $k$ increases, the value of $\sigma^2$ needed to achieve $\pe$ will decrease.  

This tradeoff is expressed by a function $\rho$. Given $\sigma^2$, let $\rho(\sigma^2, \ptarget)$ be the greatest code rate such that the decoder word-error rate $P_e(C,\sigma^2)$ is not greater than a target error rate $\ptarget$.  When $n$ is small, $P_e(C,\sigma^2)$ may be noticeably less than $\ptarget$ because $k$ is an integer. Evidently, the function $\rho$ depends on the decoding algorithm, the number of CRC bits, and the method to select the frozen bits.

For polar codes with successive cancellation decoding, $\rho$ may be found by density evolution.  Since the AMGN channel is symmetric, density evolution assuming the all-zeros codeword is sufficient. Recall that $p_j$ is the error rate for bit position $j$ assuming that positions $1$ to $j-1$ are correct, see \eqref{eqn:pi} and $\mathcal I$ is the set of information bit positions. Then the probability of word error under density evolution for polar code $C$ is:
\begin{align*}
P_e(C,\sigma^2) = 1 - \prod_{j\in \mathcal I} (1-p_j).
\end{align*}
For a fixed channel, $\rho$ is equal to the rate $R$ such that the decoder error rate $P_e(C,\sigma^2)$ under density evolution is as high as possible while satisfying $P_e(C,\sigma^2) \leq \ptarget$. 

For other decoders for which density evolution is not feasible, such as successive cancellation list decoding, $\rho$ may be found by Monte Carlo simulations. For a given number of information bits $k$ find the noise variances $\sigma^2$ which produce decoder error rates both above and below $\ptarget$.  Interpolation may helpful in improving the estimate of $\sigma^2$ which will result in $\ptarget$.

\begin{figure}[t!]
\centering		
\includegraphics[width=0.5\textwidth]{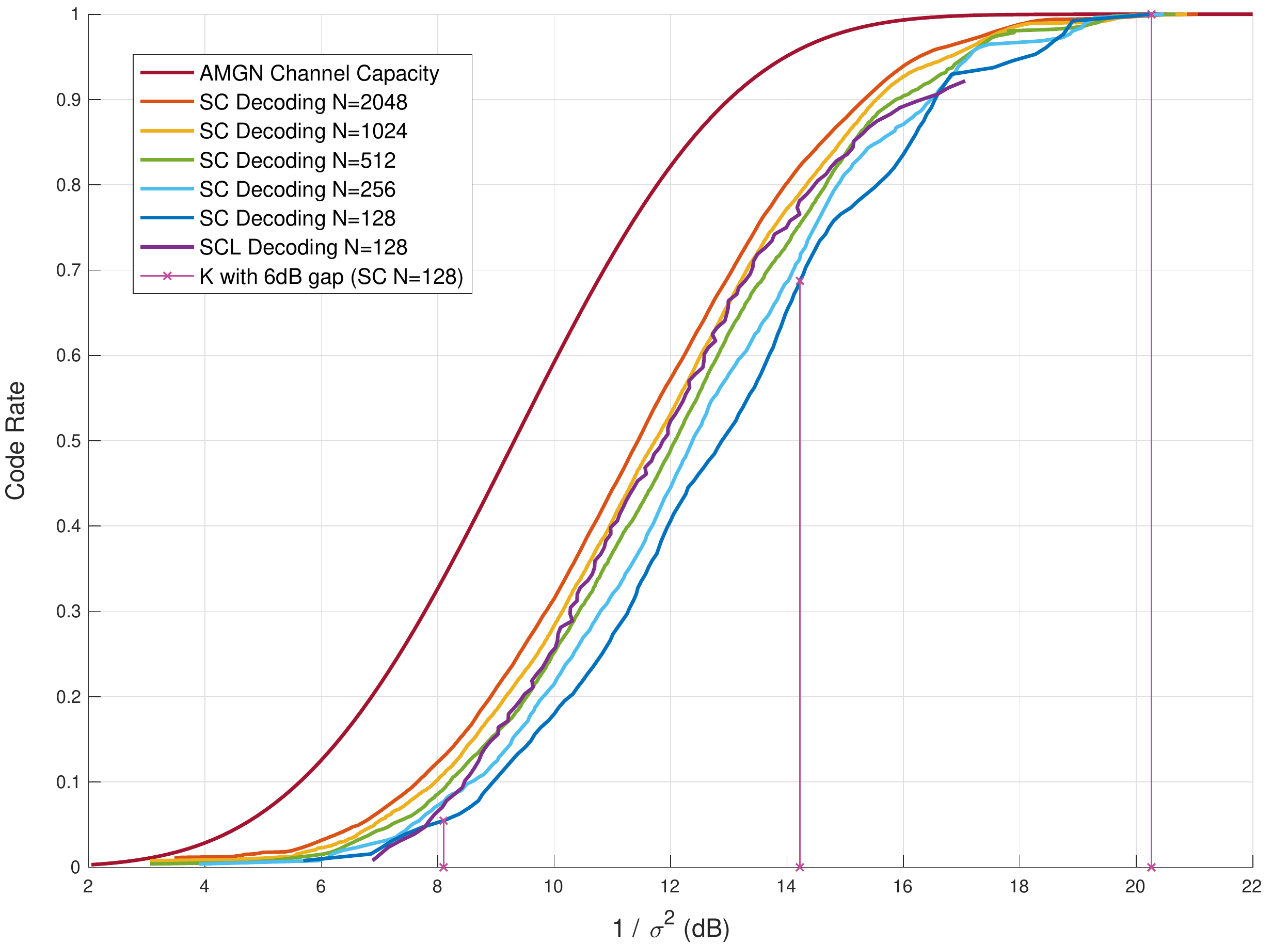}
\caption{The function $\rho(\sigma^2, \ptarget)$ is the rate $R$ for a polar code to achieve SCL decoder $\ptarget = 10^{-4}$and SC decoder $\ptarget = \frac 1 3 10^{-4}$ on a AMGN channel with noise $\sigma^2$.  The SCL decoder uses 10 CRC bits and decoder list size 8.}
	\label{fig:K_SC}
	\end{figure}

The function $\rho$ is shown in Fig.~\ref{fig:K_SC}. For successive cancellation decoding, $n=128,\ldots,2048$ are shown for a target error rate of $\ptarget = \frac 1 3 10^{-4}$.  For SCL decoding with list size 8 and 10 CRC bits, $n=128$ is shown for $\ptarget = 10^{-4}$. As expected, SC decoding requires lower rates (smaller $k$) to achieve the target $\pe$.  Also shown is the capacity of the AMGN channel  \cite[Fig.~6]{liu-com19} \cite[Fig.~7]{Wachsmann-it99}.  Interestingly, the $\rho$ curve exhibits the same S-shape characteristic of the capacity curve.

\subsection{Design of Polar Code Lattices}

To design a polar code lattice of dimension $n$, the goal is to choose $k_0,\ldots,k_{a-1}$, or equivalently rates $R_0, \ldots R_{a-1}$ such that the polar code lattice has the lowest possible VNR for a target probability of word error $\pe$. Under the equal-error probability rule, each component code $C_0, C_1, \ldots, C_a$ should have equal error rates $\ptarget$ when decoding on its respective equivalent channel.  
Each layer sees an AMGN channel as described in Section \ref{section:polarCodeLatttices}.  Let $\sigma^2_0, \sigma^2_1, \cdots$ be the AMGN noise variance for level $i = 0, 1, \cdots$.  

The $\rho$ function for a polar code of block length $n$ is used to design a polar code lattice of dimension $n$ with a designed lattice error rate of $P_e$. Following the union bound \eqref{eqn:unionBound}, choose $P_e = (a+1) \ptarget$ for an $a$-level lattice. Under the equal error probability rule, we allow $P_e(C_i, 4^{-i} \sigma^2) \approx \ptarget$, for $i=0,1,\ldots,a$.

Let $\sigma_a^2 = 4^{-a} \sigma^2$ be the noise variance for which the decoder at level $a$ achieves $\ptarget$.  Since this is the uncoded level, $k_{a} = n$:
\begin{align}
    \sigma^2_a &= \rho^{-1}(n,\ptarget) 
\end{align}    
The probability of error in level $a$ can be computed explicitly \cite[eqn.~(5)]{daSilva-it19}, and the inverse of this function gives:
\begin{align}
\sigma^2_a &=  \frac{1}{8 \cdot \left( \erfc^{-1}(1 - \sqrt[n]{1 - \ptarget})\right)^2} \textrm{,} 
\label{eqn:sigma2find}
\end{align}
where $\erfc^{-1}$ is the inverse of the complementary error function.

\begin{table}[t]
   \renewcommand{\arraystretch}{1.3}
   \caption{Polar code lattice designs under SC decoding and $\pe = 10^{-4}$.}
   \label{tab:K values}
   \centering
   \begin{tabular}{ c c c c c c}
     \hline
              & $n=64$ & $n=128$ & $n=256$ & $n=512$ & $n=1024$\\
      \hline
      $k_0$ & 1 & 7 & 24 & 68 & 178\\
      $k_1$ & 40 & 88 & 192 & 410 & 866\\
      $k_2$ & 64 & 128 & 256 & 512 & 1024\\[2pt]
      $\frac 1 {\sigma^2_a}$ &20.03 dB & 20.26 dB & 20.47 dB & 20.68 dB & 20.87 dB \\

      \hline
   \end{tabular}
\end{table}

Once $\sigma^2_a$ is fixed, use the function $\rho$ to find $R_{i}$ for $i = 0, \ldots, a-1$:
\begin{align}
    R_{i} &= \rho(4^{a-i} \sigma_a^2 ,\pe).
\end{align}
This results in nested binary codes $C_0, \ldots, C_{a-1}$ that forms a polar code lattice.

\section{Design Examples}
\label{section:Design128}

In this section, we use the proposed design method and give examples of polar code lattices with $a = 2$, evaluate their error rate and complexity. 

\subsection{Polar Code Lattices Under SC Decoding}

Under SC decoding, design a polar code lattice with $a=2$, $n=128$ and a target error rate $\pe = 10^{-4}$.
Since $a=2$, using $\ptarget = \frac 1 3 10^{-4}$ and \eqref{eqn:sigma2find} to find $\sigma_2^2 = 0.0094258$, which is $20.26$ dB.  Let $\sigs_i$ be the noise seen in the equivalent AMGN channel in level $i$, as shown in Fig.~\ref{fig:kfunction}.  Continuing the design procedure, we seek $k_1$ for which $\sigs_1$ has a 6 dB gap from $\sigs_2$, and similarly we seek $k_0$ for which $\sigs_0$ has a 6 dB gap from $\sigs_1$ (6 dB gap corresponds to the factor of 4). Since $\sigma_2^2 = 20.26 \textrm{ dB}$, from Fig.~\ref{fig:K_SC}, we obtain the design of $k_0 = 7$ at 8.26 dB and $k_1 = 88$ at 14.26 dB. 

Table~\ref{tab:K values} shows polar code lattice designs for various dimensions $n$, based on density evolution result in Fig.~\ref{fig:K_SC}. The target error rate is $\pe = 10^{-4}$.

\subsection{Polar Code Lattices Under SCL Decoding}

Successive cancellation list (SCL) decoders are considered.  SCL decoding has good performance-complexity trade-off, but CRC bits are needed. In the short block-length regime, selecting the number of CRC bits is particularly important to the performance of the SCL decoder; the number of CRC bits are chosen to balance the tradeoff between reliability and code rate \cite{Murata-isit17}.

We use the polarization weight method \cite{zhou2018weight} to allocate frozen bits, this is a channel-independent approximation method \cite{obed2019polar}.  For this method, the information bits satisfy $\mathcal I_0 \subseteq \mathcal I_1 \subseteq \cdots \subseteq \mathcal I_a$, thus the subcode condition needed by Construction D is met. 

The design to obtain $k_0$ and $k_1$ is similar to the SC design. Density evolution is not suitable for SCL decoding due to list decoding. For this design we target $P_e = 3 \cdot 10^{-4}$, so $\ptarget = 10^{-4}$ and $\sigma^2_2$ corresponds to $19.98$ dB. From Fig \ref{fig:K_SC}, we obtain the designs of $k_0 = 7$ at $7.85$ dB and $k_1 = 95$ at $13.82$ dB.

\subsection{Evaluation by Simulation}

\begin{figure}
\centering		\includegraphics[width=0.5\textwidth]{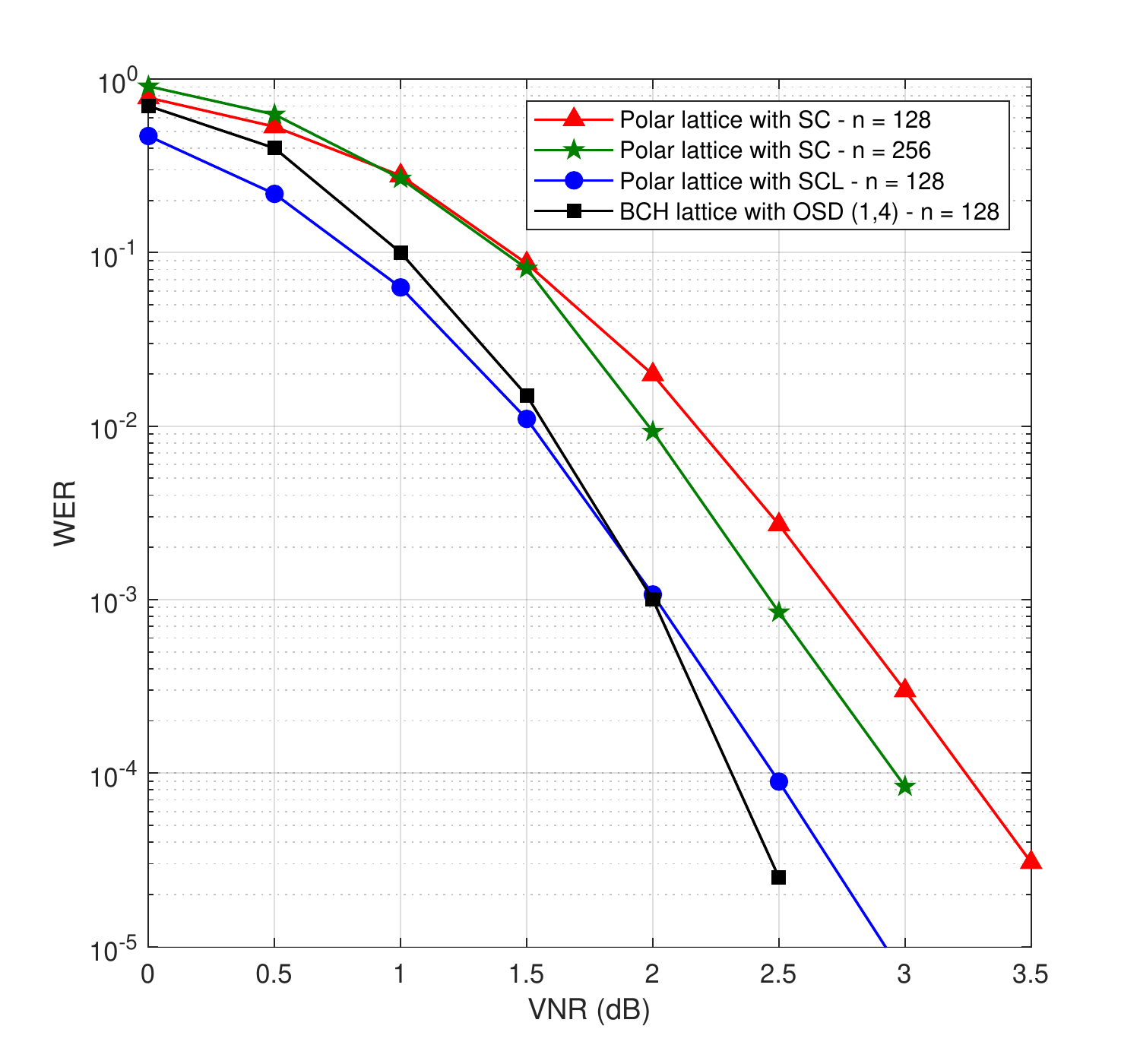}
\caption{WER on the unconstrained power channel, comparison between $n=128$ BCH code code lattice with OSD decoding \cite{matsumine2018bch} and $n=128, 256$ polar code lattices with SC decoding and SCL decoding (SCL with CRC-6 and list size 128).}
	\label{fig:result6db}
	\end{figure}

We evaluated these polar code lattices with their respective decoders by simulation, WER is shown in Fig.~\ref{fig:result6db}.  For SC decoding, the $n = 128$ polar code lattice achieves a VNR of $3.25$ dB.  The $n=256$ polar code lattice achieves a VNR of $3.0$ dB under similar conditions. For SCL decoding, the polar code lattice achieves a VNR of $2.5$ dB with 6 CRC bits and list size $L = 128$.  For reference, the WER of the $(128,120), (128,78)$ BCH code lattice with order-(1,4) OSD decoding achieves a VNR of $2.3$ dB \cite{matsumine2018bch}, all at a WER or $\pe = 10^{-4}$.

\subsection{Performance-Complexity Trade-off}

While polar code lattices comes within 0.2 dB of the BCH code lattice, OSD decoding \cite{Fossorier-it95} of BCH lattices has significantly higher complexity.  SCL decoding with list size $L$ complexity scales as $O(L n \log n)$ while order-$l$ OSD decoding complexity is proportional to $\sum_{i=0}^l {k \choose i}$.

\begin{figure}
\centering		\includegraphics[width=0.5\textwidth]{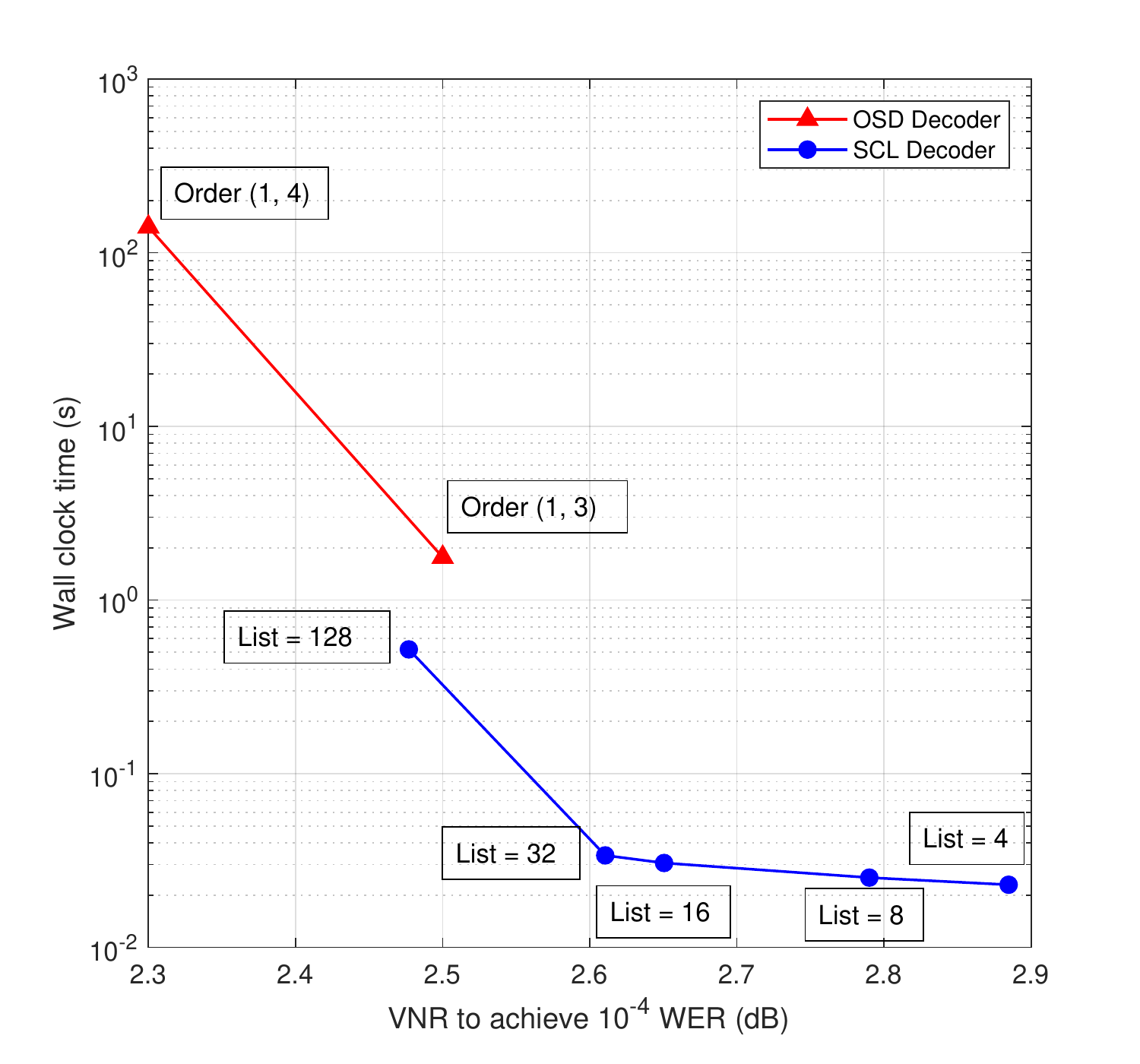}
\caption{Performance (VNR to achieve $\pe = 10^{-4}$) vs complexity (wall clock time) for OSD decoding and SCL decoding.}
	\label{fig:complex}
	\end{figure}

However, for fixed $n$, we compare complexity by evaluating the wall clock time, the complexity as measured by the average running time to decode one codeword. Fig.~\ref{fig:complex} shows the wall clock time versus VNR to achieve WER of $10^{-4}$.  We found the wall clock time of OSD decoding with order (1,3) and (1,4); the wall clock time of SCL decoding is evaluated with list sizes 4, 8, 16, 32, and 128. The result shows OSD has higher wall clock time than SCL. OSD with order (1,3) requires 1.763 seconds/codeword and order (1,4) requires 140.738 seconds/codeword. The wall clock time for SCL with list size $L = 4$ with 0.023 seconds/codeword and the wall clock time with list size $L = 128$ in 0.519 seconds/codeword.  The SCL decoding with $L = 128$ is better than OSD order (1,3) in both wall clock time and VNR, and it has lower wall clock time than OSD order (1,4), although the performance is worse for 0.177 dB gap.      

\section{Conclusion}
\label{section:Conclusion}

Polar code lattices can be designed in small dimension regime, using the equal error probability rule applied to a function $\rho$, which expresses the greatest rate which achieves a target $\pe$. To find the function $\rho$, density evolution is more efficient than Monte Carlo simulations. 

While capacity-based design \cite{liu-com19} can be used to design large-dimensional polar code lattices, this technique is not sufficient for small dimensions.  As can be seen from Fig.~\ref{fig:K_SC}, there is a significant gap between capacity and finite-length rates. 

The equal-error probability rule has already been used to design Construction D' lattices based on LDPC codes \cite{daSilva-it19}; the challenge is that Monte Carlo simulations are slow and time-consuming.  For polar codes with successive cancellation decoding, density evolution allows obtaining probability of error quickly and efficiently. 

 \bibliographystyle{IEEEtran}
 \bibliography{references}

\end{document}